# An Update on Machine Learning in Neuro-oncology Diagnostics

Thomas C Booth [1,2] [0000-0003-0984-3998]

[1] King's College London, School of Biomedical Engineering & Imaging Sciences, St. Thomas' Hospital, London. SE1 7EH. United Kingdom.

[2] Department of Neuroradiology, King's College Hospital NHS Foundation Trust, London. SE5 9RS. United Kingdom.

✉Corresponding author: tombooth@doctors.org.uk

**Abstract.** Imaging biomarkers in neuro-oncology are used for diagnosis, prognosis and treatment response monitoring. Magnetic resonance imaging is typically used throughout the patient pathway because routine structural imaging provides detailed anatomical and pathological information and advanced techniques provide additional physiological detail.

Following image feature extraction, machine learning allows accurate classification in a variety of scenarios. Machine learning also enables image feature extraction *de novo* although the low prevalence of brain tumours makes such approaches challenging.

Much research is applied to determining molecular profiles, histological tumour grade and prognosis at the time that patients first present with a brain tumour. Following treatment, differentiating a treatment response from a post-treatment related effect is clinically important and also an area of study. Most of the evidence is low level having been obtained retrospectively and in single centres.

# 1    Introduction

## 1.1    Imaging Biomarkers

A biomarker can be defined as a characteristic that is measured as an indicator of normal biological processes, pathogenic processes, or responses to an exposure or intervention, including therapeutic interventions [1]. Molecular, histologic, imaging, or physiologic characteristics are types of biomarkers. In neuro-oncology, imaging biomarkers are used for diagnosis, prognosis and treatment response monitoring.

Magnetic resonance imaging is typically used throughout the patient pathway because routine structural imaging provides detailed anatomical and pathological information and advanced techniques provide additional physiological detail. Qualitative analysis of a new intracranial mass aides diagnosis and in routine clinical practice can determine whether or not to proceed to confirmatory biopsy or resection. For example, with some basic demographic information such as the age of the patient and with some basic clinical information, such as knowledge that the mass was found incidentally whilst imaging for an unrelated condition, the qualitative routine structural imaging features of a grade 1 meningioma allow diagnosis with a high positive predictive value without the need for confirmatory biopsy. Advanced techniques allow quantitative analysis of masses which can also change management. For example, cerebral blood volume values obtained using dynamic susceptibility-weighted imaging within an area of tumour contrast enhancement, or 1H-magetic resonance spectroscopic ratios acquired from a tumour, may help determine whether a mass is of high histological grade (grade III or IV) in certain scenarios.

Some image analysis recommendations, which determine treatment response of high histological grade gliomas, have become common in the research setting and rely on simple linear metrics of simple image features, namely the product of the maximal perpendicular cross-sectional dimensions of contrast enhancing tumour [2,3].

Unlike the above biomarkers where simple imaging features are apparent to the reporting clinician, much image analysis research aims to extract underlying information from the imaging dataset to develop biomarkers that may not be readily visible. Machine learning can be applied to different phases of image analysis research which sequentially consists of pre-processing images, feature estimation (quantifying or characterizing the image), feature selection (remove noise and random error in the underlying data), classification (decision or discriminant analysis) and evaluation [4].

## 1.2 Clinical Validity

Evaluation in image analysis research initially consists of analytical validation, where accuracy and reliability of the biomarker are assessed [5]. Accuracy determines how often a test is correct in a given population (the number of true positives and true negatives divided by the number of overall tests). Clinical validation is the testing of biomarker performance in a clinical trial. Biomarkers in neuro-oncology may not be rigorously proven to be analytically or clinically valid [5]. Validation instead may attempt to use a common biomarker thereby reducing the clinical validity. For example, an attempt to validate a new imaging biomarker for treatment response monitoring may involve comparing it to a common biomarker for treatment response, such as the product of the maximal perpendicular cross-sectional dimensions of contrast enhancing tumour. However, the common biomarker itself may not be rigorously proven to be clinically valid.

This update describes several illustrative research studies with a variety of designs aimed at developing imaging biomarkers for diagnosis, prognosis and treatment response monitoring using machine learning. Different machine learning strategies used in classification in particular, as well as feature estimation and selection, are demonstrated. The extent of analytical and clinical validation is highlighted. As with the illustrative studies described here, most research studies pertaining to machine learning and neuro-oncology are pioneering but the level of evidence is low [6]. Afterall, most studies are retrospective and performed in single centres.

## 2 Diagnostic Biomarkers

## 2.1 Pre-diagnostic Biomarkers

Pre-diagnostic or risk or susceptibility biomarkers are typically clinical or molecular and occur in the absence of overt neuro-oncological disease. An example could be the discovery of a patient with Li-Fraumeni syndrome. This is a hereditary cancer syndrome due to mutations in the tumour suppressor gene p53 where patients have a susceptibility for the development of glioma. Other examples include DNA repair gene polymorphisms, single-nucleotide polymorphisms and a history of ionizing radiation [5]. Imaging has had a negligible contribution to neuro-oncological pre-diagnosis.

## 2.2 Diagnostic Biomarkers

Diagnostic biomarkers are used to detect or confirm the presence of a disease or a subtype of the disease [1]. Both histology and molecular features are now frequently combined and 1p/19q chromosome arm co-deletion status and isocitrate dehydrogenase (IDH) mutation status are routinely acquired after biopsy in accordance with the 2016 World Health Organization Classification of Tumors of the Central Nervous System [7]. There has been much research using machine learning to extract molecular information from imaging, known as radiomics. The results have been promising but prospective clinical validation is required [5].

**Example 1.** The aim of this retrospective study was to use a machine-learning algorithm to generate a model predictive of IDH mutant status in high-grade gliomas based on clinical variables and multimodal features extracted from pre-operative routine MRI [8]. True IDH mutant status was determined following biopsy using a combination of immunohistochemistry, spectrometry and sequencing. The authors suggest that knowing the pre-operative IDH mutant status might counter the limited sensitivity of immunohistochemistry and might influence the extent of tumour resection, although there is limited evidence for these assertions. Pre- and post-contrast T1-weighted, T2-weighted, and apparent diffusion coefficient map images were obtained. Whole tumour, enhancing and non-enhancing tumour volumes as well as a tumour border region were segmented. Subregions delimited by apparent diffusion coefficient thresholds within the three volumes were also segmented. Imaging descriptors including location, first and second order (textural) statistics gave 2970 extracted features. Feature selection was performed using area under the receiving-operator characteristic curve (AUC) threshold and correlation. The remaining 386 features were used to build a model predictive of IDH mutant status by applying random forest to a 90 patient training dataset. The tree depth was set to 64 with a 4096 tree upper bound limit and bootstrapping applied. Ten-fold cross validation was used. This gave 86% accuracy with an AUC of 0.88. The model was tested on a 30 patient in test dataset giving 89% accuracy and 0.92 AUC.

  Heterogeneity metrics associated with ADC-delineated segmentation were the imaging features that contributed most in predicting IDH mutant status. Despite the multiple complex imaging features such as these, patient age gave the highest predictive value of IDH mutant status demonstrating the importance of including simple, accessible information as features in radiomic analyses. Unfortunately, other simple data such as Karnofsky Performance Status, which is known to be an important covariate in multivariate analyses of glioma survival, was not included. Nonetheless, the overall approach shows that machine learning allows combinations of features to be combined to give higher accuracy than single features alone including age.

  A strength of the study is that routine imaging alone was used which makes translation to the clinic more feasible than if advanced imaging algorithms were also included. This is due to a frequent lack of standardization in many advanced imaging algorithms.

Common to most studies of diagnostic biomarkers, a limitation is that the findings relate to a single institution therefore the findings cannot be generalized elsewhere. Secondary high grade glioma were excluded, which presumably relates to exclusion of low grade gliomas that were followed up and then transformed. It is also noted that only enhancing tumours were included. Within the institution, the model can only be used within these constraints.

**Example 2.** In a similar retrospective study, a machine-learning algorithm was also applied to multimodal features extracted from pre-operative routine MRI to generate a model predictive of IDH mutant status (84 patients) [9]. In this example, grade II and III gliomas were studied and 1p/19q chromosome arm co-deletion status (67 patients) was also predicted as was grade (84 patients). Pre- and post-contrast T1-weighted, T2-weighted/FLAIR images were obtained from The Cancer Genome Atlas (TCGA)/ The Cancer Imaging Archive (TCIA) dataset. Imaging descriptors with similarities to the previous study such as location, derived from Visually Accessible Rembrandt Images (VASARI), as well as second order statistics were determined.

Second order (textural) statistics and VASARI features were independently applied to raw images that had undergone a variety of manipulations such as down-sampling or grey-scale thresholding, using different sequences to give 3360 extracted features. Feature selection was performed using logistic regression and bootstrapping was performed to maximize the area under the receiving-operator characteristic curve giving models with < 10 features. Using this methodology alone, second order statistic models performed better than VASARI models predicting IDH1 mutation status, 1p/19q co-deletion status and histological grade with AUCs of 0.86, 0.96, and 0.86, respectively. Random forest using 500 trees was then applied to combinations of clinical features and the two models of selected imaging features. IDH mutation status, 1p/19q co-deletion and histological grade were predicted with AUCs of 0.86, 0.89 and 0.78. Overall, texture played a dominant role in prediction. It is noteworthy that prediction of 1p/19q co-deletion status and grade was more accurate with logistic regression and bootstrapping methodology alone than when used as an input for random forest.

Analytical validation with a separate test dataset is required to improve analytical validity and make the findings more meaningful. However, even with further analytical validation the findings are unlikely to be translatable to the clinic as the fundamental constraint for clinical validation is that there was *a priori* knowledge that there were no grade IV gliomas in the dataset.

**Example 3.** In this small retrospective study a voxel-based unsupervised clustering method used a batch-learning self-organizing map (SOM) followed by *k*-means to determine regional histological grade from pre-operative routine MRI [10]. SOM is a neural network which can simplify features and remove outliers. *k*-means can identify features with similar patterns. Pre- and post-contrast T1-weighted and T2-weighted/FLAIR images from 36 patients with grade II-IV gliomas were processed and 161,157 extracted features underwent this two-level clustering to give clustered image maps. Segmented clustered image map regions corresponding to enhancing tumour tissue, non-enhancing tumour tissue, and oedematous tissue were described as class ratios which were used as inputs for supervised analysis. Classification was by a linear kernel support vector machine (SVM) using leave-one-out cross validation to distinguish low and high grade gliomas. The clustered image map with the optimal number of cluster classes gave an accuracy of 0.86 with 0.93 AUC. It was noted that a phenotype for high grade gliomas included high intensity of post-contrast T1-weighted and FLAIR images in contrast enhancing regions whereas a low grade phenotype showed high intensity of T2 images in these regions. Information from contrast enhancing regions alone made a large contribution to grade prediction with an accuracy of 0.82.

The method was applied prospectively to 4 patients with analysis of targeted biopsy tissue from representative classes which gave some limited evidence that the clusters gave meaningful information. It is noteworthy that no clinical parameters were used. Although this is a single centre study with a small number of patients, and without robust clinical validation, the approach to diagnostic biomarker development is an exemplar for how to minimise *a priori* knowledge.

## 3      Monitoring Biomarkers

Monitoring biomarkers are measured serially and may detect change in extent of disease, provide evidence of treatment exposure or assess safety [1]. There is an overlap with safety biomarkers which specifically determine any treatment toxicity. Monitoring blood or cerebral spinal fluid for circulating tumor cells, exosomes, and microRNAs shows promise [5]. However, imaging is particularly useful as it is non-invasive and captures the entire tumour volume and adjacent tissues and has led to recommendations to determine treatment response in trials [2,3]. Clinical validation is typically not proven. Common biomarkers are frequently used in an attempt to indirectly validate the monitoring biomarker under development.

**Example 1.** The aim of this small glioblastoma study was to use a machine-learning algorithm to differentiate progression from pseudoprogression, at the earliest time point when an enlarging MRI-enhancing lesion is seen, using T2-weighted images alone [11]. Unsupervised feature estimation was performed using principal component analysis to investigate topological descriptors of image heterogeneity called Minkowski functionals. After confounders were identified (MRI field strength) and sensitivity to field strength demonstrated, a supervised analysis was performed. Feature selection reduced Minkowski functional, first order statistical and clinical features from 32 to 7. A radial basis function kernel support vector machine gave an accuracy of 0.88 in a retrospective training dataset of 17 patients and 0.86 in a prospective test dataset of 7 patients. Although not apparent to the reporting radiologist, the T2-weighted hyperintensity phenotype of those patients with progression was heterogeneous, large and frond-like when compared to those with pseudoprogression. The pseudoprogression phenotype on T2-weighted images was shown to be a distinct entity and different from vasogenic oedema and radiation necrosis.

Additional analytical validation was performed firstly in the form of reliability testing which showed that a different operator performing segmentation achieved 100% classification concordance. Secondly, the same results using a different software package and a different operator were also obtained. Thirdly, a different feature selection method (random forest) and classifier (lasso) were used and also gave the same accuracy with 6 similar selected features.

A strength of the study is that T2-weighted images alone were used increasing the chance of translation. However, the study was performed in a single centre and, as the authors point out, the biomarker requires clinical validation in a larger multicentre test dataset.

**Example 2.** The aim of this small high grade glioma study was to use a machine-learning algorithm to differentiate progression from pseudoprogression at the earliest time point when an enlarging MRI-enhancing lesion is seen, using [18F]-fluoroethyl-L-tyrosine positron emission tomography [12]. First and second order statistics were obtained from the images of 14 patients and underwent unsupervised consensus clustering. The cumulative distribution function then determined the optimal class size. Feature selection by predictive analysis of microarrays methodology using 10-fold cross validation reduced the features from 19 to 10. One of the 3 class PET-based clusters could differentiate progression and pseudoprogression, however the results were similar to the standard analysis method using maximal tracer uptake in the tumor divided by that in normally appearing brain tissue. The small, single centre study will require more analytical and clinical validation as the authors acknowledge.

## 4     Prognostic Biomarkers

Prognostic biomarkers identify the likelihood of a clinical event, recurrence, or progression based on the natural history of the disease [1]. They are generally associated with specific outcome such as overall survival or progression-free survival. Some molecular markers are prognostic biomarkers therefore there is some overlap with diagnostic biomarkers used to predict molecular markers (including IDH mutation status and 1p/19q co-deletion status).

**Example 1.** The aim of this retrospective study was to use a machine-learning algorithm to determine overall survival using imaging features from pre-operative routine MRI in patients with glioblastoma [13]. Pre- and post-contrast T1-weighted, FLAIR, DSC and diffusion tensor imaging (DTI) images were obtained from a retrospective training dataset of 105 patients. Enhancing tumour tissue, non-enhancing tumour tissue, and oedematous tissue regions were segmented with the glioma image segmentation and registration (GLISTR) segmentation algorithm which produced imaging descriptors including location and first order statistics and limited demographic information. From > 150 features, 60 features with the best survival prediction following 10-fold cross validation were feature selected. Two linear kernel SVMs were used to classify patients as survivors or not at 6 and 18 months respectively and a combined prediction index calculated. Tenfold cross validation was used to determine the generalization accuracy of the predictive models to give an accuracy of 77% for the prediction of short/medium/long survivors. A prospective test dataset of 29 patients to gave an accuracy of 79%.

Simple data such as Karnofsky Performance Status, which is known to be an important co-variate in multivariate analyses of glioma survival, were not included. An insightful aspect of this study is that histograms were produced in order to understand the predictive features: greater age, large tumour size, increased tumour diffusivity, larger regions of T2 hypointensity and highest perfusion peak heights, were all predictive of short survival. Although the findings have a plausible biological basis, translation is limited as this was performed in a single centre.

**Example 2.** The aim of this retrospective study was to use a machine-learning algorithm to determine overall survival of patients with high grade glioma using brain tumor segmentation (BRaTS) data [14]. Pre- and post-contrast T1-weighted, T2-weighted and FLAIR images were obtained from a retrospective training dataset of 163 patients. Segmented regions including enhancing tumour tissue, non-enhancing tumour tissue, and oedematous tissue regions were manually segmented. Features were selected by simple features such as location; discrete wavelet transform first and second order statistics; histograms alone; and a convolutional neural network (CNN) which gave over 4000 deep features. The CNN, AlexNet, used in transfer learning context consisted of five convolutional layers followed by three fully connected layers, with maximum pooling layers used in between the convolution and fully connected layers.

Patients were then classified as survivors or not at 10 and 15 months respectively. SVM, *k*-nearest neighbors (KNN), linear discriminant, tree, ensemble, and logistic regression were all independently applied to each set of features. A combination of CNN deep features and a linear discriminant classifier with 5-fold cross validation gave the best predictive result with a train dataset of 91% accuracy and a test dataset of 55% accuracy. Although interesting approaches to developing a prognostic biomarker were employed including using a CNN to generate features, the low test accuracy is suggestive of overfitting.

## 5 Conclusion

Machine learning and neuro-oncology are at an early stage of development and are not ready to be incorporated into the clinic as the level of evidence is low. Integration of data in addition to imaging, including demographic, clinical and molecular markers, may lead to increasingly accurate biomarkers. Development and validation of machine learning models applied to neuro-oncology require large, well-annotated datasets, and therefore multidisciplinary and multicentre collaborations are necessary.

**Acknowledgments** This work was supported by the Wellcome/EPSRC Centre for Medical Engineering [WT 203148/Z/16/Z].